\documentclass{article}
\usepackage{amssymb,epsfig}

\newcommand{\Real}{\mathop{\mathrm{Re}}}
\newcommand{\Imag}{\mathop{\mathrm{Im}}}
\newcommand{\muB}{\mathop{\mu_\mathrm{B}}}

\begin{document}
\begin{center}
\Large\bf
Neutrino scattering on polarized electron target and
neutrino magnetic moment
\end{center}
\vskip0.5cm
\centerline{\Large\bf Timur I. Rashba}
\vskip0.5cm
\centerline{\bf E-mail: rashba@izmiran.rssi.ru}
\begin{center}
\it
The Institute of the Terrestrial Magnetism,
the Ionosphere and Radio Wave Propagation of the Russian Academy of
Sciences IZMIRAN, Troitsk, Moscow region, 142190, Russia
\end{center}
\begin{abstract}
The completed and proposed experiments
for the measurement of the neutrino magnetic moment are discussed.
To improve the sensitivity of the search for the neutrino magnetic moment
we suggest to use a polarized electron target
in the processes of neutrino (antineutrino) -- electron
scattering. It is shown that in this case the weak interaction term in the
total cross section is few times smaller comparing with
unpolarized case, but the electromagnetic term does not depend on electron
polarization.
\end{abstract}

\vskip0.5cm

The search of new bounds on the neutrino magnetic moment $\mu_\nu$ in
laboratory experiments would be very important for checking of new
Physics beyond the Standard Model (SM)
and for different astrophysical implications.  The
Resonant Spin-Flavor Precession (RSFP) scenario (Akhmedov, 1988;
Lim and Marciano, 1988)
in the case of
a non-vanishing neutrino magnetic moment $\mu_\nu\neq 0$ is still
considered as a possible solution to the Solar Neutrino Problem
(SNP) (Berezinsky, 1998).  The observations of neutron stars gave an
information about the existence of strong magnetic fields (even more than
the critical value $B_{\mathrm{c}}=m_e^2/e=4.41\cdot10^{13}$\,G) which can
interact with neutrino through its magnetic moment.  The nature of a neutrino
(Dirac or Majorana) determines the properties of neutrino magnetic
moments, so a Dirac neutrino can have diagonal and transition
(off-diagonal) moments, while Majorana neutrinos can have transition
moments only (Schechter and Valle, 1981).
All these hints claim direct laboratory
measurements of the neutrino magnetic moment.

There is the laboratory bound on the electron neutrino magnetic
moment from the experiments with reactor's neutrino
combined by Derbin (1994), $\mu_{\nu_e} <1.8\cdot10^{-10}\muB$. There
are also different astrophysical constraints that are even more stringent
then previous one. For the detailed discussion and references see
Raffelt (1996).

To probe neutrino magnetic moment at a more lower level
$\mu_{\nu_e}\sim 10^{-11}\muB$ some artificial neutrino sources of
well-known activity and spectra can be used. There are some experiments
planned with different possible artificial sources, e.g.
$^{51}$Cr (Ferrari et al., 1996), $^{90}$Sr (Mikaelyan et al.,
1998; Ianni and Montanino, 1999),
$^{147}$Pm (Kornoukhov, 1997; Barabanov et al., 1997),
$^{55}$Fe (Golubchikov et al., 1996).
The recently proposed measurement of neutrino magnetic moment
at a level better than $10^{-11}\muB$ ( $\sim 3\cdot10^{-12}\muB$)
with the use of the tritium neutrino emitter (antineutrino source with
the energy spectrum endpoint $18.6$\,keV) and a semiconductor cryodetector
is planned to reach extra low threshold about $1.1$\,eV (Trofimov et al.,
1998; Bogdanova, 2000) (see Table~1).

\begin{center}
\begin{tabular}{llll}
\hline\hline
\vphantom{|}
Reference & $\nu$($\bar\nu$) source & Detector & \hfill$\mu_\nu/\mu_B>$\\
\hline\hline
\vphantom{|}
Ferrari et al., 1996 &
$^{51}$Cr & BOREXINO & \hfill$5\cdot10^{-11}$\\
\hline
\vphantom{|}
Ianni et al., 1999 &
$^{51}$Cr, $^{90}$Sr & BOREXINO &
\hfill$6\cdot10^{-11}$\\
\hline
\vphantom{|}
Miranda et al., 1999 &
$^{51}$Cr, $^{90}$Sr & HELLAZ &
\hfill$6\cdot10^{-11}$ \\
\hline
Beda et al., 1998 & Reactor & Ge-NaI &
\hfill$3\cdot10^{-11}$  \\
\hline
TEXONO, Wong, 1999 & Reactor, Taiwan & CsI &
\hfill$3\cdot10^{-11}$\\
\hline
Kurchatov-PNPI, Kozlov, 1999 &  Krasnoyarsk reactor & Si &
\hfill$2\cdot10^{-11}$\\
\hline
MUNU, Broggini, 1999 & Bugey reactor & CF$_4$ &
\hfill$2\cdot10^{-11}$\\
\hline
Mikaelyan et al., 1998 & $^{90}$Sr & BOREXINO &
\hfill$1.5\cdot10^{-11}$\\
\hline
Golubchikov et al., 1996 &
$^{55}$Fe &
Si &
\hfill$5\cdot10^{-12}$ \\
\hline
Kornoukhov, 1997 & $^{147}$Pm & Ge, Si or NaI &
\hfill$5\cdot10^{-12}$ \\
\hline
Barabanov et al., 1997 &
$^{147}$Pm & NaI &
\hfill$3\cdot10^{-12}$ \\
\hline
Trofimov et al., 1998 &
TiT${_2}$ &
Si  &
\hfill$3\cdot10^{-12}$ \\
\hline
\end{tabular}

\begin{center}
Table 1. Proposed and planning experiments and neutrino
magnetic moments which can be measured.
\end{center}
\end{center}

We suggest to use a polarized electron target for more precise
measurements of a neutrino magnetic moment in the processes of
neutrino (antineutrino) -- electron scattering (Rashba and Semikoz, 2000).
Kinematics of such
scattering for which the recoil electron energy
$T_e  = E_2  - m_e$ should be
measured is shown in Fig.1. We have fixed $z0x$-plane based on the
initial neutrino momentum ${\vec k}_1$ directed along $z$-axis and on the
$3$-vector part ${\vec\xi}_e$ of the four-spin
$a_{\mu}  = (0, {\vec\xi}_e)$
at the rest frame of the initial electron (${\vec p}_1  = 0$) entering the
Mishel--Wightman density matrix, $\rho(p_1)  = (\hat{p}_1 + m_e)(1 +
\gamma_5\hat{a})$.

The cross section of the electron neutrino (antineutrino)
scattering off electrons
consists of two terms: weak and electromagnetic ones without interference
term that can be neglected in the massless limit $m_{\nu} \to 0$,
\begin{equation}
\left(\frac{d\sigma}{dT_ed\phi}\right) =
\left(\frac{d\sigma}{dT_ed\phi}\right)^{\mathrm{weak}} +
\left(\frac{d\sigma}{dT_ed\phi }\right)^{\mathrm{em}}~.
\label{sum}
\end{equation}
After integration over the azimuthal angle $\phi$ (see Fig.1) the weak
cross sections have the following forms:
\begin{eqnarray}
&& \left(\frac{d\sigma}{dT_e}\right)^{\mathrm{weak}}_{\nu}=\frac{2G_{F}^{2}m_{e}}{\pi}
\left[ g_{eL}^{2}\left( 1+|\vec\xi_e|\cos\theta_\xi\right) +
g_{R}^{2}\left( 1-
\frac{T_e}{\omega_{1}}\right) ^{2}\times
\right.\nonumber\\ && \left.\times
\left(1-|\vec\xi_e|\cos\theta_\xi\left(1-
\frac{m_eT_e}{\omega_1(\omega_1-T_e)}\right)\right)
-g_{eL}g_{R} \frac{m_{e}T_e}{\omega_{1}^{2}}
\left( 1+|\vec\xi_e|\cos\theta_\xi\right) \right]
\label{weak2}
\end{eqnarray}
for left-handed neutrino and
\begin{eqnarray}
&& \left(\frac{d\sigma}{dT_e}\right)^{\mathrm{weak}}_{\tilde\nu}=
\frac{2G_{F}^2m_e}{\pi} \left[ g_{R}^2\left(
1-|\vec\xi_e|\cos\theta_\xi\right) + g_{eL}^2\left(
1-\frac{T_e}{\omega_{1}}\right)^2\times
\right.\nonumber\\ && \left.
\times \left(
1+|\vec\xi_e|\cos\theta_\xi\left(1-\frac{m_eT_e}{\omega_1(\omega_1-T_e)}
\right)\right)
-g_{eL}g_{R} \frac{m_eT_e}{\omega_1^2}
\left( 1-|\vec\xi_e|\cos\theta_\xi\right) \right]
\label{weak3}
\end{eqnarray}
for right-handed antineutrino, where
$|\vec\xi_e|$ is the
value of the initial electron polarization.
Here $G_F$ is the Fermi constant;
$g_{eL} =\sin^2\theta_W + 0.5$, $g_{R} =\sin^2\theta_W$
($\sin^2\theta_W\approx0.23$) are the couplings in the SM;
$\omega_1=k_1$ is the initial neutrino energy; $\theta_\xi$ is the
angle between the neutrino momentum ${\vec k}_1$ and the electron
polarization vector $\vec{\xi}_e$.
If the new (experimental) parameter $|\vec\xi_e|\cos \theta_{\xi}$
vanishes, $|\vec\xi_e|\cos \theta_{\xi}\to 0$, our results convert
to the standard ones (Okun', 1990).

Electromagnetic terms of the neutrino
(antineutrino) scattering cross section do not depend on the electron
polarization ${\vec\xi}_e$ and have the same form for neutrinos and
antineutrinos.  If CP holds\footnote{In the Dirac neutrino case when CP
conserves the diagonal electric dipole moment is absent, $d_{\nu} = 0$.
For Majorana neutrino this form is valid if CP parities of initial and
final neutrino states are opposite hence the electric dipole moment does
not contribute, $d_{ij}= 0$ (Schechter and Valle, 1981).
For the Dirac neutrino
transition moment we should substitute $|\mu_{\nu}|^2\to
|\mu_{ij} - id_{ij}|^2$ where $\mu$ and $d$ are not separated and obey
the equality $\Real \mu\Real d + \Imag\mu\Imag d = 0$ (Raffelt, 1996).} the
electromagnetic term in Eq.~(\ref{sum}) can be written as
\begin{equation}
\left(\frac{d\sigma}{dT_e}\right)^{\mathrm{em}}=\frac{\pi\alpha ^{2}}
{m_{e}^{2}}\left( \frac{1}{T_e}-\frac{1}{\omega _{1}}\right) \frac{\left|
\mu_\nu \right| ^{2}}{\mu_{\mathrm{B}}^{2}},  \label{em}
\end{equation}
here $\muB$ is the Bohr magneton.

As well as for the case of an unpolarized target (${\vec\xi}_e= 0$) the
electromagnetic interaction via a large magnetic moment increases very
significantly in the region of small energy transfer $T_e=\omega_1 -
\omega_2 \ll m_e$, $\omega_1$ and can be comparable or greater than the weak
interaction.  This gives a possibility to determine an upper limit on the
neutrino magnetic moment or to measure it.

Weak interaction cross sections (Eqs. (\ref{weak2}) and (\ref{weak3}))
occur very sensitive both to the polarization of
electron target $|\vec\xi_e|$ and to the angle $\theta_{\xi}$ between
the neutrino momentum and the initial electron polarization.
For the known values of the SM
parameters ($g_{eL}^2\approx0.533$, $g_R^2\approx0.053$ and
$g_{eL}g_{R}\approx0.168$) one can observe from
Eqs. (\ref{weak2}) and (\ref{weak3}) a new possibility for
decreasing of the main weak term ($\sim g_{eL}^2$) in the cross section
Eq.~(\ref{sum}) choosing
large polarization values ($|\vec\xi_e|\to 1$) and with a choice of
the specific geometry ($\theta_{\xi}=\pi$). The weak cross sections
for $^{51}$Cr neutrino source and different values of the initial electron
polarization are shown in Fig.3.

In the case discussed above ($|\vec\xi_e|\cos \theta_{\xi}=-1$)
the weak interaction cross sections are proportional to $g_R^2$. It
means that target electrons (in laboratory system) that are fully
polarized in the opposite direction to the neutrino momentum will interact
with a neutrino as right chiral particles.  This can be explained by the
following way. Let's make the Lorentz boost from the laboratory system to the
ultrarelativistic one which  moves along neutrino momentum with $V\sim c=1$.
In this system initial electron is an ultrarelativistic particle
and it can be considered as a massless one. Since the modulo of electron
polarization is Lorentz-invariant, $|\vec{\xi_e}| =
\sqrt{-a_{\mu}a^{\mu}}$, and the electron momentum direction appeared in
chosen system is parallel to its polarization vector such electron is
the right-handed particle (right-helicity $=+1$) having same chirality ($+1$)
in  massless limit.

The best case of the antiparallel neutrino (antineutrino) beam $\cos
\theta_{\xi} = -1$ can be generalized for a more realistic case with an
integration over $\theta_{\xi}$ accounting for a concrete geometry of the
artificial (isotope) neutrino source placed outside of the detector with
known sizes.

Note that the polarization contribution to the ${\tilde\nu}e$-scattering
vanishes at the two points:
\begin{equation}
\label{zero}
{T_e}_1=\frac{\omega_1^2}{m_e+\omega_1}\left(1+\frac{g_R}{g_{eL}}\right)\,,\quad
{T_e}_2=\omega_1\left(1-\frac{g_R}{g_L}\right)\,.
\end{equation}
For the tritium antineutrino source
the polarization reduces (enhances) the
unpolarized part of the SM weak cross section below (above) ${T_e}_1$.
For the same source the second root ${T_e}_2$ occurs out of the kinematic
allowed region, ${T_e}_2>T_{\mathrm{max}}=2\omega_1^2/(m_e +
2{\omega_1})$. Hence in the case of the ${\tilde\nu}e$-scattering
 the low energy region $T_e< T_{e1}$
is preferable to look for an electromagnetic effect from the sum
Eq.~(\ref{sum}).

One can easily check that in the case of the
${\nu}e$-scattering ($g_{eL}\leftrightarrow g_R$) both roots
Eq.~(\ref{zero}) are out of the kinematic allowed  region and
the polarization term reduces the SM weak cross-section for any energy
$T_e$.

Let us consider the spectra of recoil electrons from an antineutrino
(neutrino) source calculated via the averaging of the differential
cross-sections over the antineutrino (neutrino)
spectrum $\rho (\omega_1)$,
\begin{eqnarray}
S^{\mathrm{weak}}_{\mathrm{free}}(T_e) &=&
\int_{{\omega_1}_{\mathrm{min}}}d{\omega_1}\frac{d\sigma^{\mathrm{weak}}
(T_e,{\omega_1})}{dT_e}\rho({\omega_1})~,
\nonumber\\
S^{\mathrm{em}}_{\mathrm{free}}(T_e) &=&
\int_{{\omega_1}_{\mathrm{min}}}d{\omega_1}\frac{d\sigma^{\mathrm{em}}
(T_e,{\omega_1})}{dT_e}\rho({\omega_1})~,
\label{freespectrum}
\end{eqnarray}
where weak and electromagnetic cross-sections are given by
Eqs.~(\ref{weak2}), (\ref{weak3}) and Eq.~(\ref{em})  correspondingly and
${\omega_1}_{\mathrm{min}} =
(T_e/2)[1 + \sqrt{1 + 2m_e/T_e}]$ is the minimal neutrino energy given by
the kinematic limit $T_e\leq T_{\mathrm{max}}$. The ratio of total recoil
electron spectra to weak one for different values of initial electron
polarization and neutrino magnetic moment for the  tritium antineutrino
source is shown in Figures 2, 6 and 7. The same plots for the $^{51}$Cr
neutrino source are shown in Figures 4 and 5.

It was found by Kopeikin et al. (1997),
 that inelastic spectra
$S_{\mathrm{in}}(q)$ which depend on the energy transfer
$q= \varepsilon_i+ T_e$ for a bound initial electron knocked out from the
$i$-shell coincide with the spectra Eq.~(\ref{freespectrum}),
$S_{\mathrm{in}}(q) = S_{\mathrm{free}}(q)\theta (q - \varepsilon_i)$, when the value $q$ is
larger than the binding energy $\varepsilon_i$, $q> \varepsilon_i$.
Really, $q$ is exactly the event energy measured in the experiment since soft
X-quanta and mainly Auger electrons emitted by an atom from which
a recoil electron was knocked out are automatically absorbed in the
detector fiducial volume or their total energy $\varepsilon_i$ adds to the
kinetic energy $T_e$ (Kopeikin et al., 1997).

For shells ordered, $i=K,L,M,N,O,\dots$, let us consider low energy transfer,
$\varepsilon_{i -1}> q\geq \varepsilon_i$, when antineutrinos
 (neutrinos) knock out outer (let us say $i= O$, main Bohr number $n=5$)
electrons only not touching all inner ones.  Moreover, since inner
electrons from filled shells with $J=L=S=0$ ($i -1 =K,L,\dots$) do not
contribute to the electron polarization one has a sense to consider only
the $\nu e$-scattering off outer $i$-electrons for which the Zeeman splitting
energy is given by
\begin{equation} E_{J,M_J}= g_J\mu_B M_JH~,
\label{Zeeman}
\end{equation}
where $g_J$ is the Lande factor and
$M_J = - J,\dots, J-1,J$ is the projection of the total spin ${\vec J}$  on the
magnetic field direction.

In order to obey such
conditions we propose to use a tritium antineutrino source
($T_{\mathrm{max}}\approx 1.26$~keV) and a cryogenic detector with the
lowest threshold ($\sim$~eV-region) for which some lower energy bins
$T_{\mathrm{th}}\leq T_{\mathrm{th}} + \Delta
T_e\leq T_{\mathrm{th}} + 2\Delta T_e,\dots$ within the interval
\begin{equation}
\label{interval}
T_{\mathrm{th}}< q= T_e + \varepsilon_i< \varepsilon_{i-1}~,
\end{equation}
can be separated. For instance, such low thresholds are considered
by Trofimov et al. (1998) for the Si-semiconductor detector without magnetic
fields. Hence the recoil electron energy $T_e$ should be lower than
{\em the difference between the binding energy of the last filled shell
$\varepsilon_{i-1}$ and outer (valence) electron energy $\varepsilon_i$},
$T_e< \varepsilon_{i-1}  - \varepsilon_i$. In such case only polarized
(outer) electrons contribute to the event spectrum to be measured.

One assumes that a low bin
width $\Delta T_e\sim T_{\mathrm{th}}\sim$ a few eV would be enough to measure the
spectrum $S_{\mathrm{in}}(q)=S_{\mathrm{free}}(q)$.

Note that we have discussed above the spectrum $S_{\mathrm{in}}(q)$ for the
ionization process $\nu + e(i)\to \nu + e$ for an individual electron on
the $i$-subshell with a free final electron. Accounting for all polarized
electrons on that outer $i$-subshell of an atom $Z$ (ion in the detector
molecule) we find  that the total sum over
subshells (Kopeikin, 1997) reduces within the low-energy
interval Eq.~(\ref{interval}) to the fraction of polarized electrons:
\begin{equation}
\frac{S_{in}(q)}{S_{\mathrm{free}}(q)}=
\frac{1}{Z}\sum_{i=K,L_\mathrm{I-III},\dots} n_i\theta(q - \varepsilon_i)\to
\frac{n_O}{Z}~, \label{fraction}
\end{equation}
where $n_i$ is the number of electrons on the $i$-subshell, $n_O$ is the
number of polarized (valence) electrons on the last unfilled, let us say,
$i=O$-shell.

The equal reduction of weak and
electromagnetic inelastic contributions $S_{in}(q)$ comparing with the
hard energy case $q\gg \varepsilon_i$ (for which $S_{\mathrm{in}}
\approx S_{\mathrm{free}}$)
leads to a calculable event statistics decrease.  Nevertheless, for low
energies Eq.~(\ref{interval}) near $T_{\mathrm{th}}$ the conserving ratio
\begin{equation}
\frac{S^{\mathrm{em}}_{\mathrm{in}}(q) +
S^{\mathrm{weak}}_{\mathrm{in}}(q)}{S^{\mathrm{weak}}_{\mathrm{in}}(q)} =
\frac{S^{\mathrm{em}}_{\mathrm{free}}(q) +
S^{\mathrm{weak}}_{\mathrm{free}}(q)}{S^{\mathrm{weak}}_{\mathrm{free}}(q)}
\label{ratio}
\end{equation}
given by Eq.~(\ref{freespectrum}) is still more sensitive to the neutrino
magnetic moment in the case of polarized outer electrons than for
unpolarized targets discussed in (Trofimov et al., 1998) (compare in Fig.~2).

The valence electrons can be fully polarized.
Really, applying realistic magnetic fields $H\gtrsim 10^4$~G to a
cryodetector one can easily reach the ratio $\muB H/T>1$ at low
temperatures $T\lesssim 0.5$~K . Known experimental
methods (Groot et al., 1965) allow to reach high electron polarization at
temperatures less than the Curie point $T_C$ when the parameter $\muB H/T$
is quite large.

For such conditions one can expect 100\,\% polarization
of outer (valence) electrons in the sum over $M_J$ entering
the electron polarization $\mid \vec{\xi}_e\mid = P$,
\begin{equation}
P = \frac{\sum_{\mid M_J\mid}[\exp (g_J\mu_B\mid M_J\mid
H/T) - \exp(- g_J\mu_B\mid M_J\mid H/T)]}{\sum_{\mid M_J\mid}
[\exp(g_J\mu_B\mid M_J\mid H/T) + \exp(- g_J\mu_B\mid M_J\mid H/T)] }~.
\label{Boltzman}
\end{equation}
Here we have used the Boltzman distribution
for independent ions (atoms) for which the fraction of the polarized
electrons with the total spin projection $M_J$ is given by the ratio
$N_{M_J}/N_0 = \exp (- g_J\mu_B M_JH/T)$ where $N_0$ is the normalization
factor not playing a role in the ratio Eq.~(\ref{Boltzman}).

Now let us turn to some experimental prospects.
Such weakly interacting Boson
gas as atomic hydrogen $H$~\footnote{Molecular
hydrogen H$_2$ is diamagnetic.} being spin-polarized in high magnetic
fields is used for possible Bose-Einstein condensation at low
temperatures (Silvera, 1995) and seems could be also used as an
ideal polarized (100~\%)
electron target for the $\nu$e-scattering.
However, we do not know how to register recoil electrons there.

The Neganov-Trofimov-Luke effect of ionization-to-heat
conversion (Neganov and Trofimov, 1981; Luke, 1988)
that was proposed for the use
of cryogenic semiconductor
detectors to measure lowest neutrino magnetic
moments (Golubchikov et al., 1996; Trofimov et al., 1998)
may be modified for such magnetic
semiconductors as ferromagnets EuS, EuO  where outer
$f$-electrons are fully polarized  along an external magnetic
field (Nagaev, 1979).

The method should be similar to Neganov and Trofimov (1981) and Luke (1988)
but has some
distinctions.  At the initial state and low temperatures $T< T_C$
($T_C\sim 14$\,K  for EuS and $T_C\sim 69$\,K for EuO) there are no
electrons (holes) in conductive (valence) zones, $n_e=n_h\sim \exp ( -
E_g/T)\approx 0$, $E_g\gg T$.  Then a neutrino hits a polarized valence
electron which overcomes the semiconductor gap $E_g$ ( $E_g = 0.9$\,eV for
EuO or $E_g = 1.5$\,eV for EuS) appearing in the conductive zone where it
could be accelerated by an external electric field of the known value
applied to the crystal to get a measurable energy $\Delta E_e$.
In the case of a simple semiconductor like Si kept at the thermostat
temperature $T$ the following electron-phonon interaction leads to the
heating of such Si-crystal for which a low lattice heat capacity $c_v\sim
(T/T_D)^3$ allows to get a large temperature jump $\Delta T\sim \Delta
E_e(T_D/T)^3$ caused by the electron energy absorption ($\sim \Delta
E_e$).

On the other hand, in a ferromagnet semiconductor the lattice heat
capacity scales as $c_v\sim (T/T_C)^{3/2}$ (Nagaev, 1979; Ziman, 1972;
White, 1970) where the
Curie temperature $T_C$ is much lower than the Debye one $T_D$, $T_C\ll
T_D$. As result the electron-magnon interaction heats crystal (through
spin-waves exited by a recoil electron) weaker than it happens
through phonons in the Si-case.  From this we conclude that new
cryogenic detectors suggested here at least should be kept at lower
temperatures while there are other possibilities that are considered now
(e.g. ferromagnet semiconductor CdCr$_2$Se$_4$ with higher $T_C\sim
110$\,K, etc).

Note that in the low energy transfer region Eq.~(\ref{interval})
the weak cross sections in the case of maximal (100\,\%)
electron polarization and for the opposite direction of
the magnetic field ${\vec H}$ with respect to the initial neutrino
(antineutrino) momentum ($|\vec\xi_e|\cos\theta_\xi=-1$) are 5 times
smaller than weak cross sections for the unpolarized case
($|\vec\xi_e|=0$).  The ratios Eq.~({\ref{ratio}) of the total recoil
electron spectra and the weak one for the tritium antineutrino scattering
off polarized electrons are shown in Figures 6 and 7.

In a real experiment the parameter $|\vec\xi_e|\cos\theta_\xi$
may in practice not equal to $-1$ because of the partially polarized
electrons $|\vec\xi_e|<1$ as well as due to the non-collinear geometry of
the experiment ($\theta_\xi\neq \pi$). E.g. the ratio of the total cross
section and the weak one for $\theta_\xi\approx 155$ degrees are shown in
Figures 4--7 where the polarization parameter $|\vec\xi_e|\cos\theta_\xi$ is
about $-0.9$ for $|\vec\xi_e|=1$.

The accuracy of $\mu_{\nu}$-calculations depends also on
experimental uncertainties including a background, an isotope source
activity, a concrete geometry of the experiment, etc. (see
e.g. Ianni and Montanino (1999)) that could be done for a future
concrete polarized cryodetector with a low threshold and is beyond the
scope of the present proposal.

Note that we have omitted radiative corrections (RC) in the
main term $g_{eL}^2(1 +\alpha f_-/\pi)(1 +|\vec\xi_e|
\cos\theta_\xi)$ (Bahcall et al., 1995) entering Eq.~(\ref{weak2}) where the
RC term $\alpha f_-/\pi$ is expected at the level $\sim 1\,\%$ in the low
energy region, or occurs at the negligible level comparing with the
strong polarization influence the weak cross-section.

I~am grateful to my co-author V.B.Semikoz for many discussions
and help in this work. I~thank M.I.Visotsky, V.A.Novikov and D.S.Gorbunov
for useful discussions during the School. I~am grateful to the Organizing
Committee of the ITEP Winter School 2000 for the pleasant and interesting
atmosphere. This work was supported by the RFBR grant 00-02-16271.

\newpage


{\large\bf References}
\begin{itemize}

\item[]
Akhmedov E.Kh., "Resonant amplification of neutrino spin rotation
in matter and the solar neutrino problem", {\it Phys. Lett.},
{\bf B 213}, 64 (1988).

\item[]
Bahcall J.N., Kamionkowski M. and Sirlin A.,
"Solar neutrinos: radiative corrections in neutrino - electron scattering
experiments", {\it Phys. Rev.}, {\bf D 51}, 6146 (1995);
e-print Archive: astro-ph/9502003.

\item[]
Barabanov I.R. et al., {\it Astropart. Phys.}, {\bf 8}, 67 (1997).

\item[]
Beda A.G., Demidova E.V., Starostin A.S., Voloshin  M.B.,
"On the feasibility of low-background Ge-NaI spectrometer for neutrino
magnetic moment measurement",
{\it Phys. Atom. Nucl.}, {\bf 61}, 66-73 (1998) [{\it Yad. Fiz.},
{\bf 61}, 72-79 (1998)].

\item[]
Berezinsky V., "Oscillation solutions to solar neutrino problem",
Talk given at 19th Texas Symposium on
Relativistic Astrophysics: Texas in Paris,, France, 14-18 Dec 1998;
e-Print Archive: hep-ph/9904259.

\item[]
Bogdanova L.N., "Studying of the neutrino properties with a superstrong
tritium source", in the Proceedings (2000).

\item[]
Broggini, C., "Status of the MUNU experiment",
{\it Nucl. Phys. Proc. Suppl.}, {\bf 70}, 188 (1999).

\item[]
Derbin A.V. et al., "Restriction on the magnetic dipole moment of reactor
neutrinos", {\it Yad. Fiz.}, {\bf 57}, 236 (1994)
[{\it Phys. Atom. Nucl.}, {\bf 57}, 222 (1994)].

\item[]
Ferrari N., Fiorentini G. and Ricci B., "The Cr-51 neutrino source and
Borexino: a desirable marriage",
{\it Phys. Lett.}, {\bf B 387}, 427 (1996).

\item[]
Golubchikov A.V., Zaimidoroga O.A., Smirnov O.Y. and Sotnikov A.P.,
"On neutrino magnetic moment measurement using artificial neutrino
source", {\it Phys. Atom. Nucl.}, {\bf 59}, 1916 (1996).

\item[]
de Groot S.R., Tolhoek H.A. and Huiskamp W.I. in
{\it Alpha-, beta- and gamma-ray spectroscopy},
edited by Siegbahn K., Vol. 3,
North-Holland Publishing Company, Amsterdam (1965).

\item[]
Ianni A. and Montanino D.,
"The Cr-51 and Sr-90 sources in BOREXINO as tool for neutrino magnetic moment
searches", {\it Astropart. Phys.}, {\bf 10}, 331 (1999).

\item[]
Kopeikin V.I., Mikaelyan L.A., Sinev V.V. and
Fayans S.A., "Reactor antineutrino-electron scattering",
{\it Yad. Fiz.}, {\bf 60} 2032 (1997).

\item[]
Kornoukhov V.N., "Laboratory source of antineutrinos on the basis of Pm-147:
conceptual design and applications",
{\it Phys. Atom. Nucl.}, {\bf 60}, 558 (1997).

\item[]
Kozlov Yu.V. et al., "Today and future neutrino experiments at Krasnoyarsk
nuclear reactor", e-Print Archive: hep-ex/9912046 (1999).

\item[]
Luke P., {\it J. Appl. Phys.}, {\bf 64}, 6858 (1988).

\item[]
Lim C.-S. and Marciano W.J., "Resonant spin - flavor precession of solar and
supernova neutrinos", {\it Phys. Rev.}, {\bf D 37}, 1368 (1988).

\item[]
Mikaelyan L.A., Sinev V.V. and Fayans S.A.,
"On a precise check of the standard model in an experiment with a Sr-90 beta
source", {\it JETP Lett.}, {\bf 67}, 453 (1998).

\item[]
Miranda O.G., Segura J., Semikoz V.B., Valle J.W.F.,
"Probing neutrino magnetic moments at underground detectors with
artificial neutrino sources",
e-Print Archive: hep-ph/9906328.

\item[]
Nagaev E.L., {\it Fizika magnitnykh poluprovodnikov}, "Nauka", Moscow, (1979),
in Russian [Nagaev E.L., {\it  Physics of magnetic semiconductors},
translated by M.~Samokhvalov, ed. by the author,
Mir Publishers, Moscow, 388 pp. (1983)].

\item[]
Neganov B.S., Trofimov V.N., {\it URSS patent} 1037771 (1981),
{\it Otkrytiya, Izobret.}, {\bf 146}, 215 (1985).

\item[]
Okun' L.B., {\it Leptons and quarks}, "Nauka", Moscow (1990).

\item[]
Raffelt G.G., {\it Stars as Laboratories for Fundamental Physics},
The University of Chicago Press (1996).

\item[]
Rashba T.I. and Semikoz V.B., "Neutrino scattering on polarized electron
target as a test of neutrino magnetic moment", {\it Phys. Lett. B}, {\bf 479},
218 (2000).

\item[]
Silvera I.F., {\it Journal of Low Temperature Physics}, {\bf 101}, 49 (1995).

\item[]
Trofimov V.N., Neganov B.S. and Yukhimchuk A.A.,
"Measurement of the neutrino magnetic moment at a level better than $10^{-11}
\mu_{B}$ with a tritium $\tilde\nu$ emitter and cryodetector (project)",
{\it Yad. Fiz.}, {\bf 61} 1373 (1998)
[{\it Phys. Atom. Nucl.}, {\bf 61} 1271 (1998)].

\item[]
Schechter J. and Valle J.W.F., "Majorana neutrinos and magnetic fields",
{\it Phys. Rev.}, {\bf D 24}, 1883 (1981);
{\bf D 25}, 283 (1982).

\item[]
White R.M., {\it Quantum Theory of Magnetism}, McGraw-Hill
Book Company, New York (1970), formula (6.64).

\item[]
Wong H.T., Li J.,
"A pilot experiment with reactor neutrinos in Taiwan",
{\it Nucl. Phys. B (Procs. Suppl.)}, {\bf 77}, 177 (1999).

\item[]
Ziman J.M., {\it Principles of the Theory of Solids}, Cambridge, The
University Press, (1972), Chapter 10, $\S~11$;

\end{itemize}

\newpage

{\large\bf Figures}

\begin{itemize}

\item[]
Figure 1. Kinematics of the neutrino scattering off the polarized
electron.

\item[]
Figure 2. The ratio of the total recoil electron spectrum and the weak
one (Eq.~\ref{ratio}) for different values of the polarization parameter
$|{\vec\xi}_e|\cos\theta_\xi=0, -1$ and for the fixed neutrino
magnetic moment $\mu_\nu=3\cdot10^{-12}\muB$ is shown as dashed and solid
lines, correspondingly. Tritium antineutrino emitter.

\item[]
Figure 3. The weak interaction term of the recoil electron spectrum
one (Eq.~\ref{ratio}) for different values of the polarization parameter
$|{\vec\xi}_e|\cos\theta_\xi=0, -0.9, -1$ is shown by dash-dotted, dashed and
solid lines correspondingly. The dotted line shows the electromagnetic term
of the recoil electron spectrum for the neutrino magnetic moment
$\mu_\nu=3\cdot10^{-12}\muB$. Neutrino source is $^{51}$Cr.

\item[]
Figure 4. The ratio of the total recoil electron spectrum and the weak
one (Eq.~\ref{ratio}) for different values of the polarization parameter
$|{\vec\xi}_e|\cos\theta_\xi=0, -0.9, -1$ and for the fixed neutrino
magnetic moment $\mu_\nu=10^{-12}\muB$ is shown by short-dashed, dashed and
solid lines, correspondingly. The zero neutrino magnetic moment case is shown
by dotted line. Neutrino source is $^{51}$Cr.

\item[]
Figure 5. The ratio of the total recoil electron spectrum and the weak
one (Eq.~\ref{ratio}) for fixed maximal value of polarization parameter
$|{\vec\xi}_e|\cos\theta_\xi=-1$ and for different neutrino
magnetic moments $\mu_\nu=0, 3\cdot10^{-13}\muB, \cdot10^{-12}\muB,
2\cdot10^{-12}\muB$ is shown by dotted, short-dashed, dashed and
solid lines, correspondingly. Neutrino source is $^{51}$Cr.

\item[]
Figure 6. The ratio of the total recoil electron spectrum and the weak
one (Eq.~\ref{ratio}) for different values of the polarization parameter
$|{\vec\xi}_e|\cos\theta_\xi=0, -0.9, -1$ and for the fixed neutrino
magnetic moment $\mu_\nu=3\cdot10^{-13}\muB$ is shown by short-dashed, dashed and
solid lines, correspondingly. The zero neutrino magnetic moment case is shown
by dotted line. Tritium antineutrino emitter.

\item[]
Figure 7. The ratio of the total recoil electron spectrum and the weak
one (Eq.~\ref{ratio}) for fixed maximal value of polarization parameter
$|{\vec\xi}_e|\cos\theta_\xi=-1$ and for different neutrino
magnetic moments $\mu_\nu=0, 10^{-13}\muB, 3\cdot10^{-13}\muB,
10^{-12}\muB$ is shown by dotted, short-dashed, dashed and
solid lines, correspondingly. Tritium antineutrino emitter.

\end{itemize}

\begin{figure}[p]
\begin{center}
\psfig{file=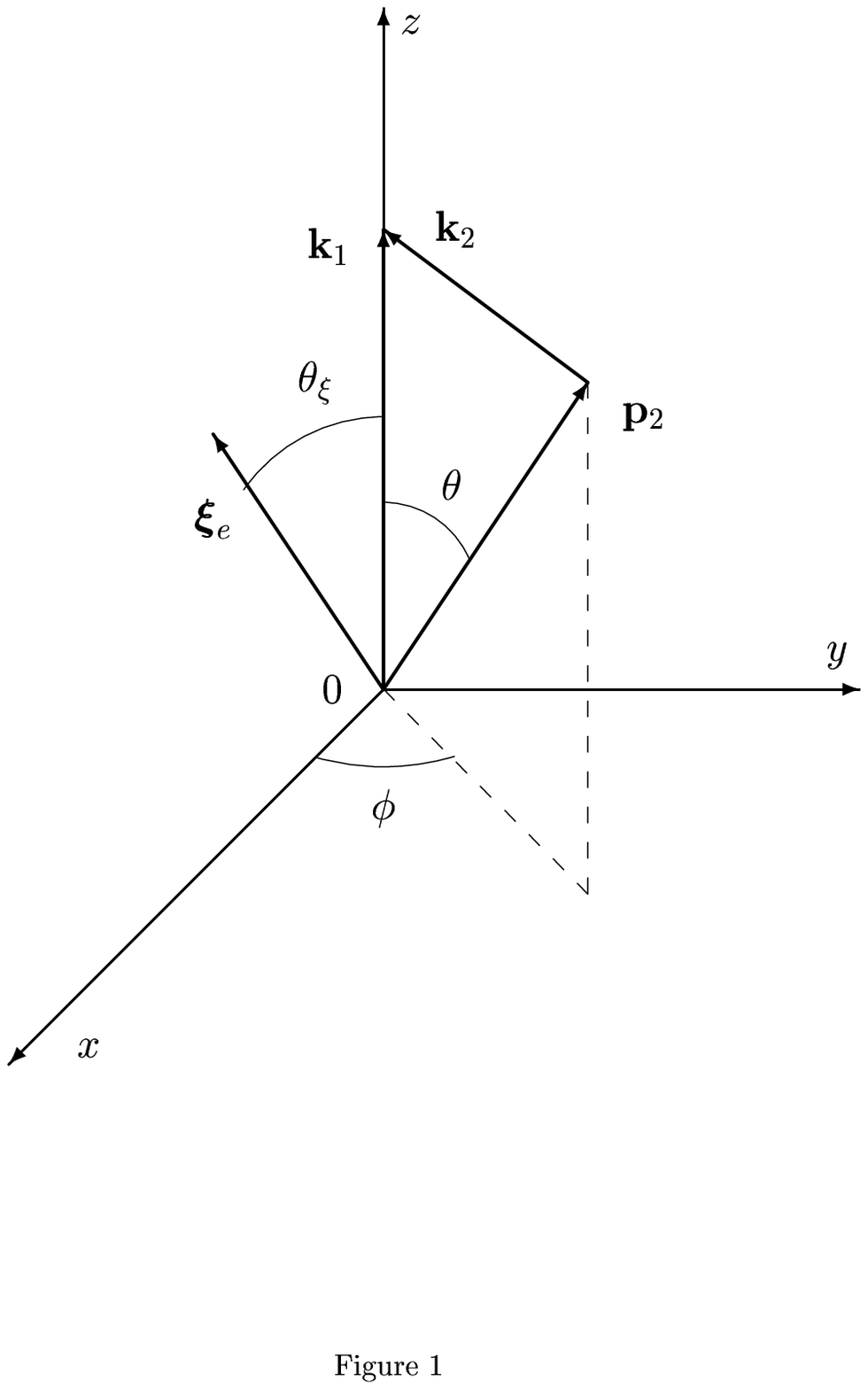} 
\end{center}
\end{figure}

\begin{figure}[p]
\begin{center}
\psfig{file=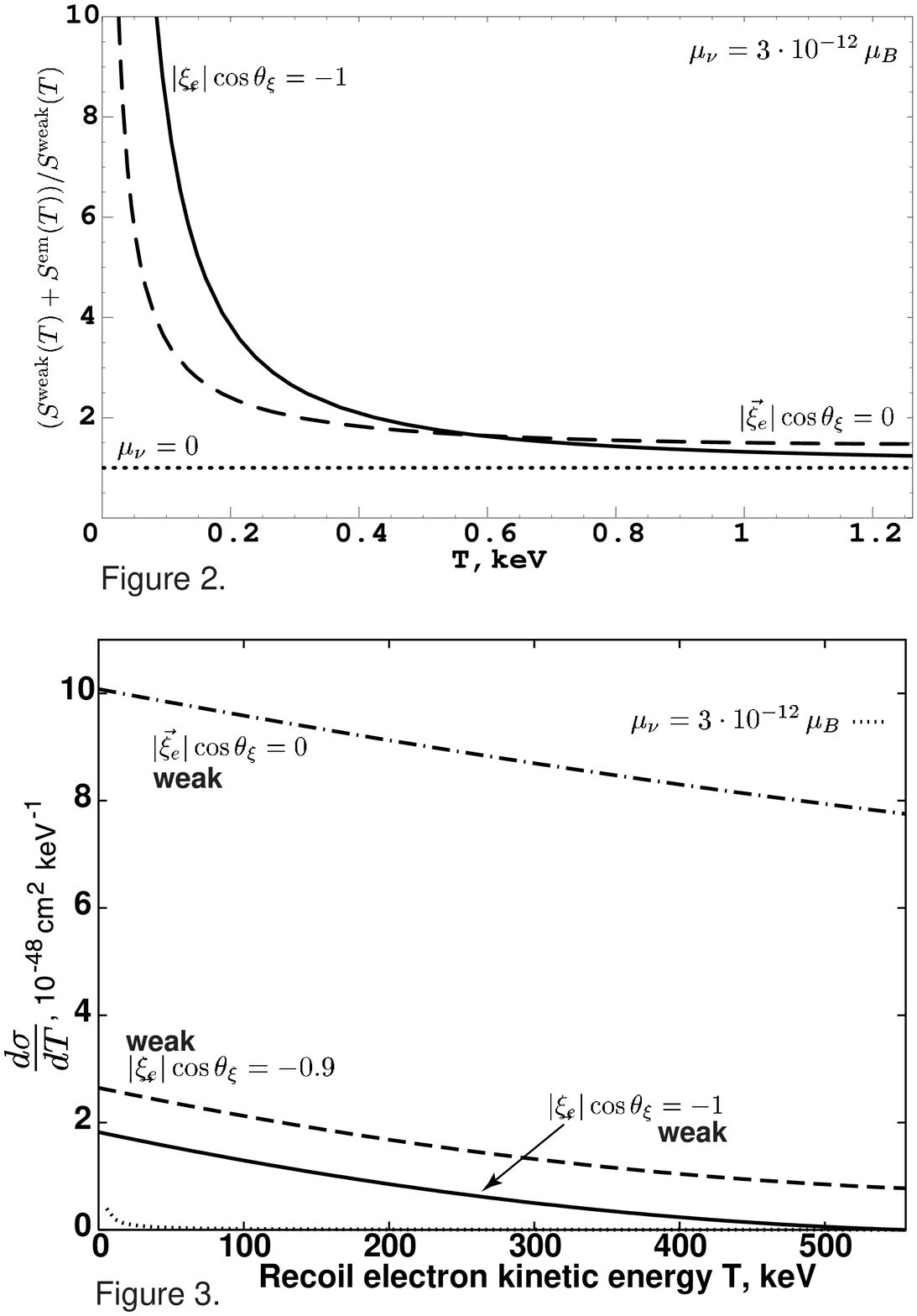,height=250mm,width=170mm,angle=0}
\end{center}
\end{figure}

\begin{figure}[p]
\begin{center}
\psfig{file=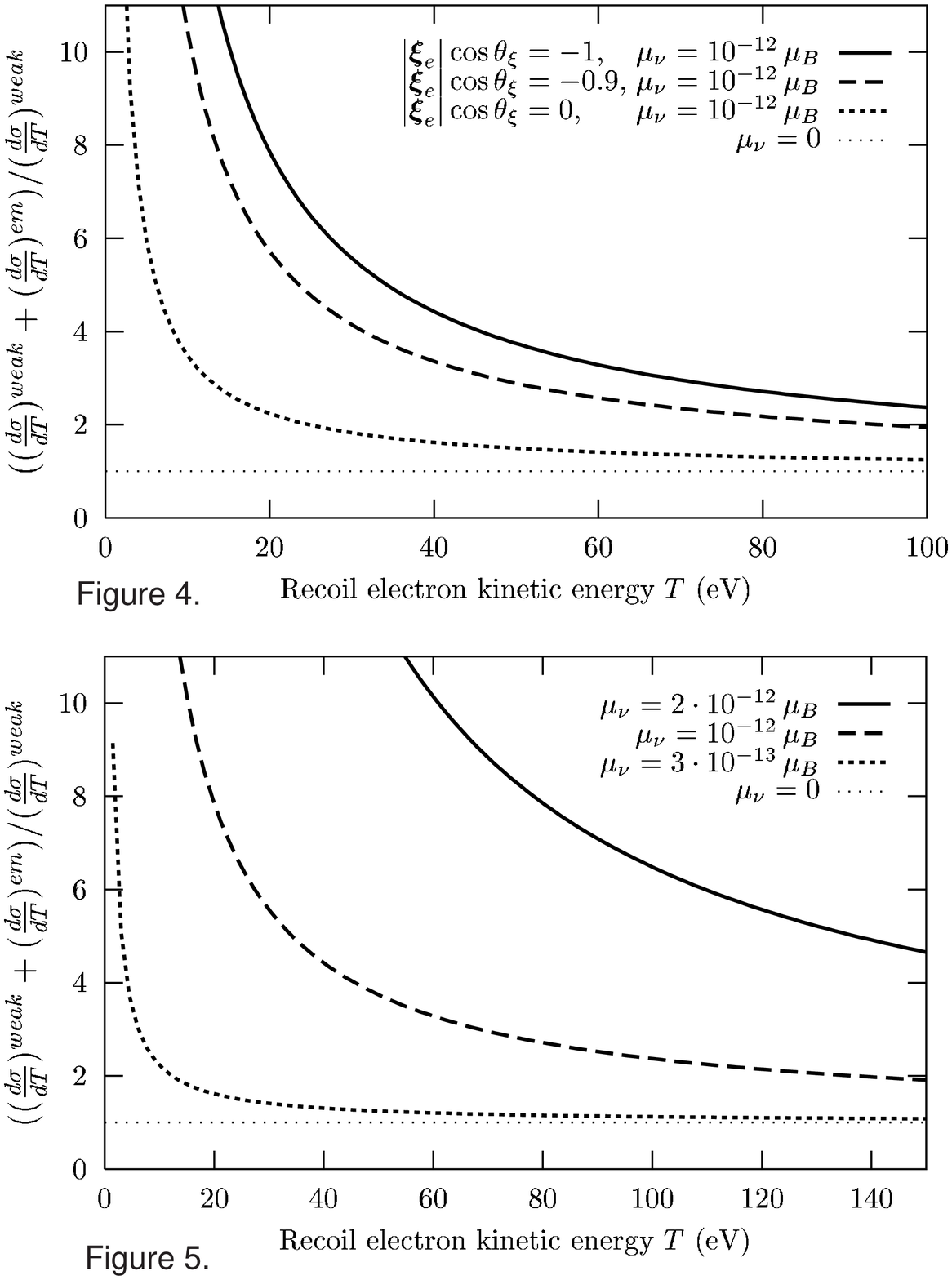,height=250mm,width=170mm,angle=0}
\end{center}
\end{figure}

\begin{figure}[p]
\begin{center}
\psfig{file=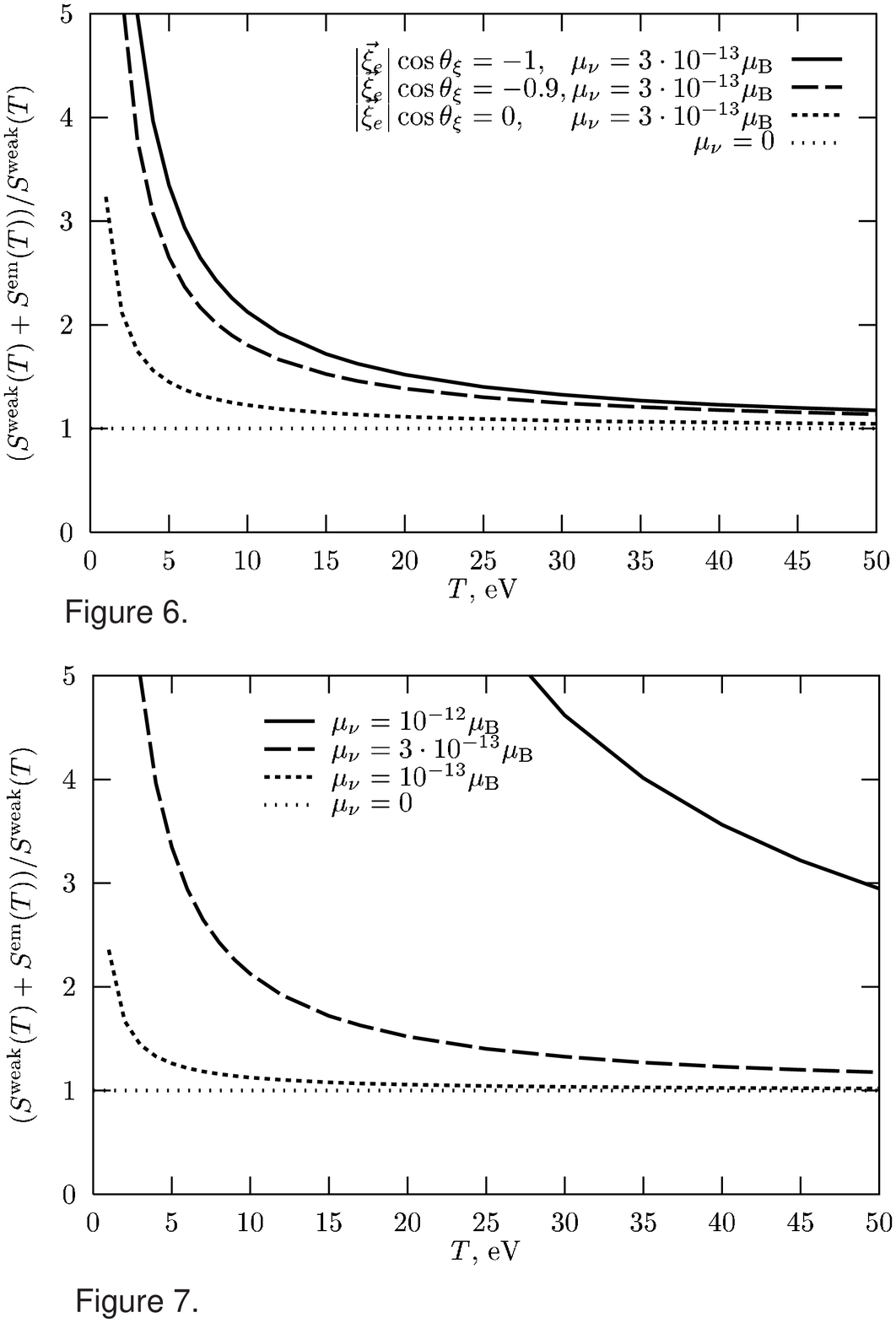,height=250mm,width=170mm,angle=0}
\end{center}
\end{figure}

\end{document}